\documentstyle[12pt]{article}

\newcommand{\rf}[1]{(\ref{#1})}
\newcommand{\bea}{\begin{eqnarray}}
\newcommand{\eea}{\end{eqnarray}}

\newcommand{\e}{\mbox{e}}

\newcommand{\del}{\delta}
\newcommand{\Del}{\Delta}

\newcommand{\ra}{\rangle}
\newcommand{\la}{\langle}

\newcommand{\mi}{\!-\!}
\newcommand{\equ}{\!=\!}
\newcommand{\pl}{\!+\!}

\def\void{}
\def\labelmark{}

\newenvironment{formula}[1]{\def\labelname{#1}
\ifx\void\labelname\def\junk{\begin{displaymath}}
\else\def\junk{\begin{equation}\label{\labelname}}\fi\junk}%
{\ifx\void\labelname\def\junk{\end{displaymath}}
\else\def\junk{\end{equation}}\fi\junk\labelmark\def\labelname{}}

{\ifx\void\labelname\def\junk{\end{array}\end{displaymath}}
\else\def\junk{\end{array}\right.\end{equation}}
\fi\junk\labelmark\def\labelname{}\def\junk{}
}

\newcommand{\beq}{\begin{formula}}
\newcommand{\eeq}{\end{formula}}
\newcommand{\beqv}{\begin{formula}{}}

\begin{document}

\hfill AEI-2000-1

\hfill NBI-HE-2000-3


\hfill  Feb 7, 2000

\begin{center}
\vspace{24pt}
{ \Large \bf A non-perturbative Lorentzian path integral\\
\vspace{12pt}
 for gravity}

\vspace{30pt}

{\sl J. Ambj\o rn}$\,^{a,}$
{\sl J. Jurkiewicz}$\,^{b,}$ and
{\sl R. Loll}$\,^{c,}$

\vspace{24pt}

$^a$~The Niels Bohr Institute, \\
Blegdamsvej 17, DK-2100 Copenhagen \O , Denmark\\
{\it email: ambjorn@nbi.dk}

\vspace{10pt}
$^b$~Institute of Physics,\\
Jagellonian University,\\
Reymonta 4, PL 30-059 Krakow, Poland\\
{\it email: jurkiewi@thrisc.if.uj.edu.pl}

\vspace{10pt}
$^c$~Albert-Einstein-Institut,\\
Max-Planck-Institut f\"{u}r Gravitationsphysik,\\
Am M\"uhlenberg 1, D-14476 Golm, Germany\\
{\it email: loll@aei-potsdam.mpg.de}

\vspace{48pt}

\end{center}


\begin{center}
{\bf Abstract}
\end{center}

A well-defined regularized path integral
for Lorentzian quantum gravity in three and four dimensions
is constructed, given in terms of
a sum over dynamically triangulated causal space-times. 
Each Lorentzian geometry and its associated action 
have a unique Wick rotation to the Euclidean sector. 
All space-time histories possess a distinguished notion
of a discrete proper time. For finite lattice volume,
the associated transfer matrix is self-adjoint and bounded.
The reflection positivity of the model ensures the existence of
a well-defined Hamiltonian. The degenerate geo\-metric phases
found previously in dynamically triangulated Euclidean
gravity are not present.  
The phase structure of the new Lorentzian quantum gravity
model can be readily investigated by both analytic and numerical 
methods.

\vspace{12pt}
\noindent


\newpage


In spite of numerous efforts, we still have not found a consistent
theory of quantum gravity which would enable us to describe local
quantum phenomena in the presence of strong gravitational fields.
Moreover, the entanglement of technical problems
with more fundamental issues concerning the structure of such a
theory often makes it difficult to pinpoint why any particular 
quantization program has not been successful.
There seems to be a broad consensus that a correct non-perturbative
treatment should involve in an essential way 
the full ``space of all metrics'' (as
opposed to linearized perturbations around flat space) and
the diffeomorphism group, i.e. the invariance group of the classical
gravitational action,
\beq{action}
S [g_{\mu\nu}]=\frac{k}{2} \int d^dx\; \sqrt{-\det g}\, (R-2\Lambda),
\qquad d=4,
\eeq  
with $k^{-1}=8 \pi G_{\rm Newton}$ and the cosmological constant
$\Lambda$.

One approach that does not rely on the existence of supersymmetry
tries to define the theory by means of a non-perturbative path 
integral. The aim is not to evaluate this in a stationary-phase
approximation, but as a genuine ``sum over all geometries''.
Since even the pure gravity theory in spite of its large invariance 
group possesses local field degrees of freedom, such a sum must
be regularized to make it meaningful. The two most popular approaches,
quantum Regge calculus and dynamical triangulations \cite{review}, 
both employ simplicial discretizations of space-time, on which
then the behaviour of metric and matter fields is studied. 
One drawback of these (mainly numerical) investigations is that 
so far they have been conducted only for {\it Euclidean} space-time
metrics $g_{\mu\nu}^{eucl}$ instead of the physical, Lorentzian
metrics. This is motivated by the analogy with non-perturbative
Euclidean (lattice) field theories on a fixed, flat background,
whose results can under suitable conditions be ``Wick-rotated"
to their Minkowskian counterparts. The amplitudes $\exp iS$
are substituted in the Euclidean path integral by the real weights
$\exp (-S^{eucl})$ for each configuration. {\it Real} weight factors
are mandatory for the applicability of standard Monte Carlo 
techniques and, more generally, for the convergence of the
state sums.  

Unfortunately, it is not at all clear how to relate path
integrals over Euclidean geometries to those over Lorentzian
ones. This has to do with the fact that in a generally covariant
field theory like gravity the metric is a dynamical quantity,
and not part of the fixed background structure. For general metric
configurations, there is no preferred notion of ``time", and hence
no obvious prescription of how to Wick-rotate.  
 
One might worry that with a discretization of space-time the
diffeomorphism invariance of the continuum theory is irretrievably
lost. However, the example of two-dimensional Euclidean gravity 
theories provides evidence 
that this is not necessarily so. There, one can choose a 
conformal gauge in the continuum formulation, and obtain an
effective action by a Faddeev-Popov construction. Physical quantities
computed in this approach coincide with those computed using an
intermediate discretization in the form of so-called
dynamical triangulations. In this latter approach one approximates
the sum over all metrics modulo diffeomorphisms by a sum over all 
possible equilateral triangulations of a given topological manifold.
Local geometric degrees of freedom are given by the geodesic
lengths (all equal to some constant $a$) of the triangle edges 
and by deficit angles around vertices. Since different triangulations
correspond to different geometries, the set-up has 
no residual gauge invariance, and it is straightforward to
define (regularized versions of)
diffeomorphism-invariant correlation functions. In the continuum
limit, as the diffeomorphism-invariant cutoff $a$ is taken to zero,
the sum over triangulations gives rise to an effective measure on
the space of geometries (whose explicit functional form is not known). 

This makes the regularization by dynamical triangulations, which is
also applicable in higher dimensions, an
appealing method for investigating quantum gravity theories.
It is a further advantage that the formalism is amenable to
numerical simulations, which have been conducted extensively
in dimensions 2, 3 and 4. Alas, in $d\equ 3,4$, and for Euclidean signature, 
an interesting continuum limit has not been found. 
This seems to be related to the dominance of degenerate
geometries. At small bare gravitational coupling $G_{\rm Newton}$,
the dominant configurations are branched polymers (or ``stacked
spheres") with
Hausdorff dimension $d_H\equ 2$, whereas at large $G_{\rm Newton}$
the geometries ``condense" around one or several singular
links or vertices with a very high coordination number,
resulting in a very large $d_H$. These extreme geometric
phases and the first-order phase transition separating them
are qualitatively well described by mean field calculations
\cite{mean,crump}.
Another unsatisfactory aspect of the Euclidean model 
(as well as other discretized models of gravity) is 
our inability to rotate back to Lorentzian space-time.

In order to tackle
these problems, two of us have recently constructed a Lorentz\-ian
version of the dynamically triangulated gravitational path 
integral in {\it two} dimen\-sions \cite{al}. The basic building blocks
are triangles with one space-like and two time-like edges.
The individual Lorentzian geometries are 
glued together from such triangles in a way that satisfies certain
causality requirements. The model turns out to be exactly
soluble and its associated continuum theory lies in a new
universality class of 2d gravity models distinct from the usual
Euclidean Liouville gravity. One central lesson from the
two-dimensional example is that the causality conditions
imposed on the Lorentzian model act as a ``regulator" for
the geometry. Most importantly, they suppress changes in the
spatial topology, that is, branching of baby universes is
not allowed. As a result, the effective quantum geometry in
the Lorentzian case is smoother and in some senses 
better behaved: a) in spite of large
fluctuations of the geometry, its Hausdorff dimension has the
canonical value $d_{H}\equ 2$, unlike in Euclidean gravity,
which has a fractal dimension $d_{H}\equ 4$; b) in spite of
a strong interaction between matter and gravity when
the system is coupled to Ising spins, the combined system
remains consistent even beyond the $c\equ 1$ barrier, unlike
what happens in Euclidean gravity \cite{aal1,aal2}.

Motivated by these results, we have constructed a 
discretized Lorentzian
path integral for gravity in 3 and 4 space-time dimensions.
Unlike in two dimensions, the action is no longer trivial,
and the Wick rotation problem must be solved.
Also the geometries themselves are more involved,
and the geometry of the $(d\mi 1)$-dimensional spatial
slices is no longer described by just a single variable.
We have succeeded in constructing a model with the 
following properties:
\begin{itemize}
\item[(a)] Lorentzian space-time geometries (histories) are
obtained by causally gluing sets of Lorentzian building 
blocks, i.e. $d$-dimensional simplices with simple length
assignments;

\item[(b)] all histories have a preferred discrete notion of 
proper time $t$; $t$ counts the number of evolution 
steps of a transfer matrix between adjacent spatial
slices, the latter given by $(d\mi 1)$-dimensional 
triangulations of equilateral Euclidean simplices;

\item[(c)] for a fixed space-time volume $N_d$ 
(= number of simplices of 
top-dimension), both the Euclidean and the Lorentzian 
discretized gravity actions are bounded from above and below;

\item[(d)] the number of possible triangulations is exponentially
bounded as a function of the space-time lattice volume;

\item[(e)] each Lorentzian discrete geometry can be ``Wick-rotated"
to a Euclidean one, defined on the same (topological) 
triangulation;

\item[(f)] at the level of the discretized action, the ``Wick rotation"
is achieved by an analytical continuation in the dimensionless
ratio $\alpha\equ -l^{2}_{\rm time}/l^{2}_{\rm space}$ 
of the squared time- and space-like link length; 
for $\alpha\equ -1$ one obtains
the usual Euclidean action of dynamically triangulated gravity;

\item[(g)] for finite lattice volume, the model is (site) 
reflection-positive, and the transfer matrix is symmetric and bounded, 
ensuring the existence of a well-defined Hamiltonian operator;

\item[(h)] the extreme phases of degenerate geometries found in
the Euclidean models cannot be realized in the Lorentzian
case.

\end{itemize}

For the sake of definiteness and simplicity, we will concentrate
mostly on the three-dimensional case. The discussion carries
over virtually unchanged to $d\equ 4$, the details of which will be given
elsewhere \cite{ajl}. (Obviously, the corresponding continuum
theories will be very different, one describing a topological
quantum field theory, and the other a field theory of interacting
gravitons.)
The classical continuum action is simply 
eq.\ (\ref{action}), with $d\equ 3$. Each discrete Lorentzian 
space-time will be given
by a sequence of two-dimensional compact spatial slices
of fixed topology, which for simplicity we take to 
be that of a two-sphere.
Each slice carries an integer ``time" label $t$,
so that the space-time topology is $I\times S^2$, with $I$ some real
interval. 
The metric data will be encoded by triangulating this underlying
space by three-dimensional simplices with definite edge 
length assignments. There are two types of edges: 
``space-like" ones (of length squared $l^2\equ a^{2} >0$, with the
lattice spacing $a>0$) which are entirely 
contained in a slice $t\equ const.$,
and ``time-like" ones (of length squared $l^2\equ -\alpha a^{2}<0$) 
which start at some slice $t$ and end at the
next slice $t\pl 1$. This implies that all lattice vertices are located
at integer $t$. 

A metric space-time is built up by ``filling in"
for all times the three-dimensional sandwiches between $t$ and $t\pl 1$.
We only consider regular gluings which lead to simplicial
{\it manifolds}.
Our basic building blocks are given by three types of 
Lorentzian tetrahedra, 
\begin{itemize}
\item[(1)] 3-to-1 tetrahedra (three vertices contained in slice $t$ and
one vertex in slice $t+1$): they have three space- and three time-like
edges; their number in the sandwich $[t,t\pl 1]$ 
will be denoted by $N_{31}(t)$;
\item[(2)] 1-to-3 tetrahedra: the same as above, but upside-down; the
tip of the tetrahedron is at $t$ and its base lies in the slice
$t+1$; notation $N_{13}(t)$;
\item[(3)] 2-to-2 tetrahedra: one edge (and therefore two vertices)
at each $t$ and $t+1$; they have two space- and four time-like edges;
notation $N_{22}(t)$.
\end{itemize}
 
Each of these triangulated space-times carries a discrete causal
structure obtained by giving each time-like link an orientation
in the positive $t$-direction. Then, two lattice vertices are causally
related if there is a sequence of positively oriented links 
connecting the two.  

The discretized form of the Lorentzian action (\ref{action}) is
obtained from Regge's prescription for simplicial manifolds 
\cite{regge}, see \cite{ajl} for details.
The action is written as a function of the deficit angles around the
one-dimensional edges and of the three-dimensional volumes,
which in turn can be expressed as functions of the squared edge
lengths of the fundamental building blocks. 
The contribution to the action from a single sandwich $[t,t\pl 1]$
is
\beq{sandwich}
\Delta S_\alpha (t)=4\pi a k\sqrt{\alpha}+(N_{31}(t)+N_{13}(t)) 
(a k K_{1} -a^{3}\lambda L_{1})
+N_{22}(t) (a k K_{2}- a^{3} \lambda L_{2}) ,
\eeq
where
\bea
K_{1}(\alpha)&=&\pi\sqrt{\alpha}-3\, {\rm arcsinh}\,
\frac{1}{\sqrt{3}\ \sqrt{4\alpha +1}} 
-3 \sqrt{\alpha}\arccos{\frac{2\alpha +1}{4 \alpha +1}},
\nonumber\\
K_{2}(\alpha)&=&2\pi \sqrt{\alpha} +2 \, {\rm arcsinh}\, 
\frac{2\sqrt{2}\ \sqrt{2\alpha +1}}{4 \alpha +1} 
-4\sqrt{\alpha} 
\arccos{\frac{-1}{4\alpha +1}},\nonumber\\
L_{1}(\alpha)&=&\frac{\sqrt{3\alpha +1}}{12},\qquad
L_{2}(\alpha)=\frac{\sqrt{2\alpha +1}}{6\sqrt{2}}.\label{kalpha}
\eea
Note that the sandwich action (\ref{sandwich}) already contains 
appropriate boundary contributions, such that $S$ is additive under the
gluing of contiguous slices.
In relation (\ref{sandwich}), we have used the rescaled cosmological
constant, $\lambda =k\Lambda$. 

At each time $t$ the physical states $|g\ra$ are parametrized by 
piece-wise linear geometries, given by unlabelled triangulations $g$
of $S^2$ in terms of equilateral Euclidean triangles. 
For a finite spatial volume $N$ (counting the triangles in a spatial slice),
the number of states is exponentially bounded as a function of $N$ and 
the vectors $|g\rangle $, defined to be orthogonal, span a 
finite-dimensional  Hilbert space ${\cal H}_{N}$. The transfer matrix
$\hat T_N$ will act on the Hilbert space
\beq{hilbert}
H^{(N)} := \bigoplus_{i=N_{\rm min}}^N {\cal H}_i,
\eeq
and the states $|g\ra$ will be normalized according to
\beq{norm}
\la g_1 | g_2 \ra = \frac{1}{C_{g_1}} \del_{g_1,g_2}, ~~~~
\sum_g C_g \; |g\ra \la g | = \hat{1}.
\eeq
The symmetry factor $C_g$ is the order of the automorphism group of the 
two-di\-men\-sional triangulation $g$, which for large triangulations 
is almost always equal to 1. 
With each step  $\Del t \equ 1$ we can now associate a transfer matrix
${\hat T}_N$ describing the evolution of the system from $t$ to $t\pl 1$,
with matrix elements  
\beq{transfer}
\langle g_2|\hat T_N(\alpha ) |g_1\rangle\equiv G_\alpha (g_1,g_2;1)
=\sum_{{\rm sandwich} (g_1\rightarrow g_2)} \e^{i \Delta S_\alpha}.
\eeq
The sum is taken over all distinct interpolating three-dimensional
triangulations of the ``sandwich" with boundary geometries $g_1$
and $g_2$, each with a spatial volume $\leq N$. 
The propagator $G_N(g_1,g_2;t)$ for arbitrary time intervals
$t$ is obtained by iterating the transfer matrix $t$ times,
\beq{prop}
G_N(g_1,g_2;t)=\langle g_2|\hat T_N^t|g_1\rangle,
\eeq
and satisfies the semigroup property
\beq{iter}
G_N(g_1,g_2;t_1+t_2)=\sum_g C_g\ G_N(g_2,g;t_2)\  G_N(g,g_1;t_1).
\eeq
The infinite-volume limit is obtained by letting $N \to \infty$ in 
eqs.\ \rf{transfer}-\rf{iter}.
  
A brief remark is in order on our notion of ``time": the label $t$ is
to be thought of as the discretized analogue of proper time,
as experienced by an idealized collection of freely falling
observers. We do not claim that this is a distinguished notion
of time for pure quantum gravity, but it is {\it a} possible choice,
in the present case suggested by our regularization. 
Note that in {\it continuum} formulations the proper time gauge is not
usually considered, because it is a gauge choice that --
considered for arbitrary geometries -- goes bad in an arbitrarily
short time. This problem does not occur in the discrete case: 
by construction we only sum over
space-time geometries for which there is a globally
well-defined (discrete) ``proper time". 

The action $S$ associated with an entire space-time $S^1\times S^2$
of length $t$ in time-direction is obtained by summing expression
(\ref{sandwich}) over all $t'=1,2,\ldots t$ and identifying 
the two boundaries. 
The result can be expressed as a function of three
geometric ``bulk" variables, for example, the total number of vertices 
$N_0$, the total number of tetrahedra $N_3$ and $t$, 
\bea
S_{\alpha}(N_{0},N_{3},t)&=&
N_{0}\left( 4 ak (K_{1}-K_{2}) -4 a^{3}\lambda (L_{1}-L_{2})\right)
+N_{3} \left( ak K_{2}-a^{3}\lambda L_{2}\right)
\nonumber\\
&& +t\left(4 ak (\pi\sqrt{\alpha}-2 
(K_{1}-K_{2}))+8 a^{3}\lambda (L_{1}-L_{2})\right). 
\label{totact}
\eea
Because of the well-known inequality $N_0 \leq (N_3\pl 10)/3$, valid
for all closed three-dimensional simplicial manifolds, this implies
the boundedness of the discretized Lorentzian action at fixed
three-volume. This result is analogous to what happens in
Euclidean dynamical triangulations. We write the partition
function as
\beq{partition}
Z_\alpha (k,\lambda,t)=\sum_{T\in{\cal T}_t(S^1\times S^2)} 
\e^{i S_\alpha (N_0(T),N_3(T),t(T))},
\eeq
with ${\cal T}_t(S^1\times S^2)$ denoting the set of all Lorentzian
triangulations of $S^1\times S^2$ of length $t$.
It will turn out that a
necessary condition for the existence of a meaningful
continuum limit is the exponential boundedness of the number
of possible triangulations as a function of the space-time
volume $N_3$: only if the growth is at most exponential in
$N_3$, can this divergence potentially be counterbalanced by a 
cosmological constant term exponentially damped in $N_3$.
In our case, exponential boundedness follows trivially
from the same property for Euclidean triangulations
(where it has been proven rigorously for $d\equ 3,4$
\cite{carfora,acm}), since
the Lorentzian space-times form a subset of the former. 
Note that the
convergence of the partition function implies the absence of
divergent ``conformal modes".

As it stands, the sum (\ref{partition}) over complex amplitudes 
has little chance of converging, due to the contributions of
an infinite number of triangulations with arbitrarily large volume 
$N_{3}$. In order to make it well-defined, one must perform a Wick
rotation, just as in ordinary quantum field theory. 
Thanks to the presence of a distinguished global time variable in
our model, we can associate a unique Euclidean triangulated space-time
with every Lorentzian history contributing in (\ref{totact}).
It is obtained by taking the {\it same} topological triangulation
and changing the squared lengths of all time-like edges from
$-\alpha a^{2}$ (Lorentzian) to $\alpha a^{2}$ (Euclidean),
leaving the space-like edges unchanged. We can then
use Regge's prescription for calculating the (real)
Euclidean action $S_{\alpha}^{\rm E}(N_{0},N_{3},t)$ associated 
with the resulting Euclidean metric space-time (where $\alpha$
is always taken to be positive). 
After some algebra one verifies that by a suitable analytic
continuation in the complex $\alpha$-plane from positive to
negative real $\alpha$, the Euclidean and Lorentzian actions
are related by
\beq{wick}
S_{-\alpha}(N_{0},N_{3},t)=iS^{\rm E}_{\alpha}(N_{0},N_{3},t),
\eeq
for $\alpha >\frac{1}{2}$. This latter inequality has its origin
in the
triangle inequality for Euclidean triangles: $S^{\rm E}$ is real
only for $\alpha \geq \frac{1}{2}$, and the limit 
$\alpha \equ\frac{1}{2}$ corresponds to the degenerate case of
totally collapsed triangles. Moreover, $\alpha\equ -1$ is
the only point on the real axis in which the coefficient of
$t$ in the Lorentzian action (\ref{totact}) vanishes
(this corresponds to $\alpha\equ 1$ in eq.\ (\ref{wick})). In this case 
one rederives the familiar expression employed in equilateral 
Euclidean dynamical triangulations, namely,
\beq{euclact}
\frac{1}{i}S_{-1}\equiv S_{1}^{\rm E}=
-ak (2\pi N_{1}-6 N_{3} \arccos\frac{1}{3})+
a^{3}\lambda N_{3}\frac{1}{6\sqrt{2}}.
\eeq
Our strategy for evaluating the partition function is now clear: for
any choice of $\alpha >\frac{1}{2}$, continue (\ref{totact})
to $-\alpha$, so that
\beq{wick2}
\sum_{T\in {\cal T}_t(S^1\times S^2)} 
\e^{i 
S_{\alpha}(N_0,N_3,t)}\;\stackrel{\alpha\to
-\alpha}{\longrightarrow}
\sum_{T\in {\cal T}_t(S^1\times S^2)} 
\e^{- 
S_{\alpha}^{\rm E}(N_0,N_3,t)}.
\eeq
Because of the exponential boundedness, the Wick-rotated
Euclidean state sum in (\ref{wick2}) {\it is} now convergent
for suitable choices of the bare couplings $k$ and $\lambda$.
We can therefore proceed in two ways: either attempt to 
perform the sum analytically, by solving the combinatorics
of possible causal gluings of the tetrahedral building
blocks (as has been done in $d\equ 2$ \cite{al}), 
or use Monte-Carlo methods to simulate the system
at finite volume. Once the continuum limit has been performed,
we can rotate back to Lorentzian signature by an analytic
continuation of the continuum proper time $T$ (which in the
case of canonical scaling is of the form $T\equ a t$; not
to be confused with the triangulation $T$) to $iT$. 
If we are only interested in vacuum expectation values of
time-independent observables and the properties of the Hamiltonian, 
we do not need to perform the Wick rotation explicitly, just as in
usual Euclidean quantum field theory. 

Let us now establish some properties of the discrete real
transfer matrix $\hat T\equiv \hat T (\alpha\equ \mi 1)$ 
of our model that are necessary for the 
existence of a well-defined Hamiltonian, defined as 
$\hat h:= -\frac{1}{2 a} \hat T^{2}$. These will be
useful in any proof of the existence of a self-adjoint {\it continuum}
Hamiltonian $\hat H$. It is difficult to imagine 
boundedness and positivity arising in the limit from regularized
models that do not have these properties. 

We will show that ${\hat T}_N$ is symmetric,
bounded for finite spatial volume $N$, and that 
the two-step transfer matrix $\hat T^{2}$ is positive.
Symmetry is obvious by inspection. 
The sandwich action (\ref{sandwich}) is
symmetric under the exchange of in- and out-states (corresponding
to $N_{31}\leftrightarrow N_{13}$). So is (\ref{transfer}), because
the counting of interpolating states does not depend on
which of the geometries $g_1$, $g_2$ defines the in-state, say.
The boundedness of ${\hat T}_N$ for finite spatial volume $N$ follows
from the finite dimensionality of the Hilbert space ${\cal H}^{(N)}$
it acts on and the fact that there is only a {\it finite}
number of possibilities to interpolate between two given spatial
triangulations $g_1$ and $g_2$ in one step. 
Positivity of the two-step transfer matrix, ${\hat T}_N^{2} \geq 0$,
follows from the reflection positivity of our model under
reflection with respect to planes of constant integer-$t$ \cite{ajl} 
(for regular lattices, this property is also 
referred to as site-reflection positivity, c.f. \cite{momu}).
In order to be able to define a Hamiltonian as
\beq{Ithehamiltonian}
\hat h_{N}:= -\frac{1}{2 a} \hat T^{2}_{N},
\eeq
we must make sure that the square of the transfer matrix is 
{\it strictly} positive, ${\hat T}_N^{2} > 0$. We do not expect
that $\hat T_{N}$ has any zero-eigenvectors, because this would
imply the existence of a ``hidden'' symmetry of the discretized
theory. It may of course happen that there are ``accidental''
zero-eigenvectors for certain values of $N$. In this case, we will
adopt the standard procedure of removing the subspace ${\cal N}^{(N)}$
spanned by these vectors from the Hilbert space, resulting in a
physical Hilbert space given by the quotient 
$H_{ph}^{(N)}=H^{(N)}/{\cal N}^{(N)}$.

It should be emphasized that although the summation in the path
integral is performed in the ``Euclidean sector" of the theory,
our construction is not {\it a priori} related to any path integral 
for Euclidean gravity proper. The point, already made in the
two-dimensional case \cite{al}, is that we sum only over a selected 
class of geometries, which are equipped with a causal structure. 
Such a restriction
incorporates the Lorentzian nature of gravity and has no
analogue in Euclidean gravity. We therefore expect our Lorentzian
statistical mechanics model to have a totally
different phase structure from that of Euclidean dynamical
triangulations. This expectation is corroborated by an
analysis of the ``extreme phases" of Lorentzian
quantum gravity, to determine which configurations dominate the
path integral 
\beq{euclpart}
Z_\alpha^{\rm E} (k,\lambda,t)=\sum_{T\in{\cal T}_t(S^1\times S^2)} 
\e^{- S^{\rm E}_\alpha },
\eeq
for either very small or very large inverse Newton's constant
$k>0$. In order to make a direct comparison with the 
Euclidean analysis \cite{gab,crump}, we set without loss of
generality $\alpha\equ 1$ in eq.\ (\ref{euclpart}) and rewrite the
Euclidean action (\ref{euclact}) as
\beq{euclact1}
S_{1}^{\rm E}=k_3 N_3 -k_1 N_1,
\eeq
with the couplings
\beq{couplings}
k_1=2\pi a k,\qquad
k_3=6\ a k \arccos\frac{1}{3}+
a^{3}\lambda \frac{1}{6\sqrt{2}}.
\eeq
In the thermodynamic limit $N_3\to\infty$, and assuming a scaling
behaviour such that $t/N_3\to 0$, one derives kinematical bounds
on the ratio of links and tetrahedra, $\xi := N_1/N_3$, namely,
\beq{ratio}
1\leq\xi\leq \frac{5}{4}.
\eeq
This is to be contrasted with the analogous result in the
Euclidean case, where $1\leq\xi\leq \frac{4}{3}$. It implies
that the branched-polymer (or ``stacked-sphere") configurations,
which are precisely characterized by $\xi =\frac{4}{3}$, and
which dominate the Euclidean state sum at large $k_1$, cannot
be realized in the Lorentzian setting.  
The opposite extreme, at small $k_1$, is associated with the
saturation of the inequality 
\beq{crumpled}
N_1\leq \frac{1}{2} N_0 (N_0 -1),
\eeq
and in the Euclidean theory goes by the name of ``crumpled
phase". At equality, every vertex is connected to every other
vertex, corresponding to a manifold with a very large Hausdorff 
dimension.
Again, it is impossible to come anywhere near this phase in
the continuum limit of the Lorentzian model. Instead of 
(\ref{crumpled}), we have now separate relations for the numbers 
$N_1^{({\rm sl})}$ and $N_1^{({\rm tl})}$ of space- and time-like
edges,
\beq{inequa}
N_1^{({\rm sl})}=\sum_t (3 N_0(t)-6)=3 N_0-6 t,\qquad
N_1^{({\rm tl})}\leq \sum_t N_0(t) N_0(t+1).
\eeq
Assuming canonical scaling, the right-hand side of 
ineq.\ (\ref{crumpled}) behaves like (length)$^6$, whereas the
second relation in (\ref{inequa}) scales only like
(length)$^5$. 

We conclude that the phase structure of Lorentzian gravity 
must be very different from that of the Euclidean theory.
In particular, the extreme branched-polymer and crumpled
configurations are not allowed in the Lorentzian theory. This
is another example of causal structure acting as a
``regulator" of geometry. It also raises the hope that the
mechanism governing the phase transition will be different,
and potentially lead to a non-trivial continuum theory,
in three as well as in four dimensions.

\vspace{.4cm}
\noindent {\it Acknowledgements.} J.A.\ acknowledges the support of 
MaPhySto -- Center for Mathematical Physics 
and Stochastics -- financed by the National Danish Research Foundation.
J.J.\ acknowledges partial support through KBN grants no. 2P03B 019 17
and 2P03B 998 14.


\begin{thebibliography}{xx}

\bibitem{review} for recent reviews, see 
J.\ Ambj\o rn, J.\ Jurkiewicz and R.\ Loll:
Lorentzian and Euclidean Quantum Gravity -- 
Analytical and Numerical Results, in: {\sl M-Theory and Quantum Geometry},
eds. L. Thorlacius and T. Jonsson, NATO Science Series (Kluwer
Academic Publishers, 2000) 382-449, hep-th/0001124;
R.\ Loll: Discrete approaches to quantum gra\-vi\-ty in four dimensions,
{\it Living Reviews in Relativity} {\bf 13} (1998), 
http://www.livingreviews.org, gr-qc/9805049.

\bibitem{mean} 
P.\ Bialas, Z.\ Burda, B.\ Petersson and J.\ Tabaczek:
Appearance of mother universe and singular vertices in random  geometries,
{\it Nucl.\ Phys.\ B} {\bf 495} (1997) 463-476 [hep-lat/9608030];
P.\ Bialas, Z.\ Burda and D.\ Johnston:
Condensation in the backgammon model,
{\em Nucl.\ Phys.\ B}\ {\bf 493} (1997) 505-516.

\bibitem{crump}
J.\ Ambjorn, M.\ Carfora, D.\ Gabrielli and A.\ Marzuoli:
Crumpled triangulations and critical points in 4D simplicial 
quantum  gravity,
{\em Nucl.\ Phys.\ B} 542 (1999) 349-394,
hep-lat/9806035.

\bibitem{al} J.\ Ambj\o rn and R.\ Loll:
Non-perturbative Lorentzian quantum gravity, causality and 
topology change,
{\em Nucl.\ Phys.\ B}\ 536 (1998) 407-434, hep-th/9805108.

\bibitem{aal1} J.\ Ambj\o rn, K.N.\ Anagnostopoulos and R.\ Loll: 
A new perspective on matter coupling in 2d quantum gravity, 
{\it Phys.\ Rev.\ D} 60 (1999) 104035, hep-th/9904012.

\bibitem{aal2} J.\ Ambj\o rn, K.N.\ Anagnostopoulos and R.\ Loll: 
Crossing the c$=$1 barrier in 2d Lorentzian quantum gravity,
{\it Phys.\ Rev.\ D} 61 (2000) 044010, hep-lat/9909129.

\bibitem{ajl} J.\ Ambj\o rn, J.\ Jurkiewicz and R.\ Loll:
Dynamically triangulating Lorentzian quantum gravity,
{\it preprint} Golm AEI-2001-049.

\bibitem{regge}T.\ Regge: 
General relativity without coordinates,
{\em Nuovo\ Cim.\ A}\ 19 (1961) 558-571;
R.\ Sorkin: 
Time-evolution in Regge calculus,
{\em Phys.\ Rev.\ D}\ 12 (1975) 385-396; Err. {\em ibid.}\ 23 (1981) 565.

\bibitem{carfora}M.\ Carfora and A.\ Marzuoli:
Entropy estimates for simplicial quantum gravity,
{\em J.\ Geom.\ Phys.}\ {\bf 16} (1995) 99-119; 
Holonomy and entropy estimates for dynamically triangulated manifolds,
{\em J.\ Math.\ Phys.}\ {\bf 36} (1995) 6353-6376.

\bibitem{acm}J.\ Ambj\o rn, M.\ Carfora and A.\ Marzuoli: 
{\em The geometry of dynamical triangulations}, 
Lecture Notes in Physics, New Series, m50, Springer, Berlin, 1997,
hep-th/9612069. 

\bibitem{momu}I.\ Montvay and G.\ M\"unster:
{\em Quantum fields on a lattice},
Cambridge University Press, Cambridge, UK, 1994.

\bibitem{gab} D.\ Gabrielli:
Polymeric phase of simplicial quantum gravity,
{\em Phys.\ Lett.\ B} 421 (1998) 79-85,
hep-lat/9710055.



\end{thebibliography}
\end{document}